\documentclass{aip-cp}

\usepackage[numbers]{natbib}
\usepackage{rotating}
\usepackage{graphicx}
\newcommand{\ket}[1]{\left\vert #1\,\right\rangle}
\newcommand{\bra}[1]{\left\langle #1\,\right\vert}
\newcommand{\braket}[2]{\left\langle #1\,\vert #2\,\right\rangle}

\begin{document}

\title{ Localized Thermal States}

\author[aff1,aff2]{Fausto Borgonovi}
\eaddress{fausto.borgonovi@unicatt.it}
\author[aff3,aff4]{Felix M. Izrailev}
\eaddress{felix.izrailev@gmail.com}
\affil[aff1]{Dipartimento di Matematica e Fisica and Interdisciplinary Laboratories for Advanced Materials Physics,
  Universit\`a Cattolica, via Musei 41, 25121 Brescia, Italy}
\affil[aff2]{Istituto Nazionale di Fisica Nucleare,  Sezione di Pavia, 
  via Bassi 6, I-27100,  Pavia, Italy}
\affil[aff3]{ Instituto de F\'{i}sica, Benem\'{e}rita Universidad Aut\'{o}noma
  de Puebla, Apartado Postal J-48, Puebla 72570, Mexico}
\affil[aff4]{Dept. of Physics and Astronomy, Michigan State University, E. Lansing, Michigan 48824-1321, USA}

\maketitle

\begin{abstract}
It is believed that thermalization in closed systems of interacting particles can occur only when the eigenstates are fully delocalized and chaotic in the preferential (unperturbed) basis of the total Hamiltonian. Here we demonstrate that at variance with  this common belief the typical situation in the systems with two-body inter-particle interaction is much more complicated and allows to treat as thermal even eigenstates that are not fully delocalized. Using   a semi-analytical approach we establish the conditions for the emergence of such thermal states in a model of randomly interacting bosons. Our numerical data show an excellent correspondence with the predicted properties of {\it localized thermal eigenstates}.

\end{abstract}

\section{INTRODUCTION}
In the last decade a large number of papers have been devoted to the onset of thermalization in closed systems of interacting fermions and bosons (see, for example, Refs.\cite{reviews,our2016} and references therein). An increase of interest in this problem is mainly due to new experimental achievements \cite{exp} and remarkable results have been obtained allowing to understand the many-body physics in view of various theoretical predictions \cite{zele,flam,deutsch,sred,kota}. Although the term {\it thermalization} is not uniquely defined, the main property of thermalization is assumed to be a possibility of a proper statistical description of the relaxation of a system to a steady state distribution, in connection with the properties of the energy spectra and eigenstates. A specific question is the so-called quench dynamics according to which one can predict the time dependence of various observables. 

The problem of thermalization in isolated systems was a basic one in order to establish a link between the classical mechanics described by deterministic dynamical equations of motion, and the statistical mechanics which is essentially a probabilistic theory. For a long time it was believed that this problem, formulated as the {\it foundation of statistical mechanics}, could find its solution with the use of  {\it ergodic hypothesis} according to which an individual classical trajectory occupies densely the whole phase space defined by the energy of a system. However, after many attempts to prove this hypothesis under general conditions, it was finally understood that the ergodicity is an exceptional property of physical systems, and can occur in quite specific systems such as dispersive billiards \cite{sinai}.

The problem of ergodicity in dynamical many-body systems was one of the initial interests of E.Fermi. After an unsuccessful attempt to prove this hypothesis in 1923 \cite{fermi}, he retained interest in this problem for many years, and in 1953, together with J.Pasta and S.Ulam, he initiated the numerical study of  thermalization in a one-dimensional linear chain of classical particles, with an additional non-linear interaction between them \cite{FPU}. It was expected that when the number of  particles is large (in the numerical experiments it was 32 and 64), the non-linearity will lead to the spread of the energy initially concentrated in one of few linear modes, over all other modes. Unexpectedly, the typical system behavior   was very different, demonstrating a kind of quasi-periodicity of the motion. This effect was found to be stable with respect to the change of model parameters and for different types of non-linearity.  

During a decade many attempts have been made in order to explain the absence of thermalization in the Fermi-Pasta-Ulam (FPU) model (see, for example, the review \cite{BI05} and references therein). Finally, in Ref.~\cite{IC66}  the condition for the onset of strong chaos was analytically obtained and the latter was claimed to be considered as the mechanism for thermalization. This condition essentially depends on the number of particles, strength of non-linearity, total energy and initial conditions; specifically  it was shown that for the model parameters used in Ref.~\cite{FPU} the criterion for the onset of chaos was not fulfilled. Few year later, the analytical predictions have been confirmed numerically \cite{chir1}, with a clear manifestation of  thermalization by choosing proper values of the model parameters.

The FPU-problem and its resolution is quite instructive for the understanding of  thermalization in quantum systems of interacting particles. Indeed, first one should expect (and this was manifested in many studies \cite{zele,sred,flam,our2012,our2016}) that the emergence of  quantum thermalization is related to the onset of quantum chaos. Second, a large number of interacting particles does not necessarily lead to thermalization, and one should also take into account other control parameters such as the total system energy, the type of particles (Fermi or Bose) and the degree of the inter-particle interaction. Moreover, when discussing the quench dynamics, it is very important to take into account that the dynamics strongly depend on the choice of the initially chosen wave packets. Below we demonstrate how all these parameters influence the onset of quantum chaos, and therefore, the thermalization in a model of randomly interacting Bose-particles.

As was noted by Landau and Lifshitz in their {\it Statistical Mechanics} book, a complete description of quantum statistical mechanics can be done either on the level of individual states or by means of the Gibbs distribution. This statement may be considered as a kind of quantum ergodicity, however it does not give the answer to the question under which conditions this fact holds in physical reality. One of the first numerical observations of this fact was reported in Ref.\cite{JS85} for a model of interacting spins in a magnetic field. In particular, it was shown that the mean value of the magnetization does not depend on whether it is computed for individual eigenstates or by  averaging over a number of eigenstates with close energies.  Also, a quite impressive manifestation of the emergence of the Fermi-Dirac distribution on the level of a single eigenstate was reported in Ref.\cite{gold} when computing, from the first principles, the eigenstates of highly excited atoms of gold. An important contribution to the problem of thermalization in terms of the conventional statistical distributions (Boltzmann, Fermi-Dirac and Bose-Einstein) has been done in Ref.\cite{sred}. It was analytically shown that under the assumption of completely chaotic eigenstates these distributions can emerge on the level of single eigenstates. Another important result is the suggestion to use band random matrices for the study of the thermalization in closed systems of interacting particles \cite{deutsch}. 

The goal of this paper is to show that for isolated systems, even in the limit case of a completely random two-body interaction, the eigenstates typically occupy only a part of the energy space defined by the Hamiltonian. However, they can be treated as chaotic ones provided the threshold of quantum chaos is achieved. Such eigenstates, presented in the unperturbed basis, consist of very many components and the correlations between them can be neglected. We demonstrate that although the eigenstates are globally localized in the unperturbed basis, one can speak of thermalization occurring on the level of individual eigenstates (see, also, \cite{our2017}). With this word we mean that the occupation number distribution for a small system of interacting particles follows the   Bose-Einstein distribution   when a proper renormalization of the energy of these eigenstates is done.

\section{THE MODEL}
 
We consider $N$ identical quantum particles occupying $M$ single-particle levels. It is assumed
that the particles follow the Bose statistics, therefore, the Hilbert space generated by the
many-particle basis states has the dimension
\begin{equation}
  N_H= \frac{(N+M-1)!}{N!(M-1)!} \, .
\end{equation}
The Hamiltonian of the model considered here can be separated into two parts,
\begin{equation}
  H = H_0 + H_I \, ,
\end{equation}
where $H_0$ stands for non-interacting particles and $H_I$ describes the interaction between particles. In the many-particle basis $H_0$ can be written as follows,
\begin{equation}
  H_0 = \sum_{s=1}^M \epsilon_s \, a^\dag_s a_s \, ,
\end{equation}
where $\epsilon_s$ are the energies of single-particle levels and $a^\dag_s$, $a_s$ are the
creation and annihilation operators. The single-particle spectrum has been chosen as random,
with the $M$ values of $\epsilon_s$ uniformly distributed between $0$ and $M$, so that the mean (single-particle) level spacing is \mbox{$d =1$.} We have also considered the case of  regular picked-fence single-particle levels (not shown here)  and the main results were found to be pretty much the same. For the interaction part $H_I$ we consider a \emph{two-body random interaction} (TBRI),
\begin{equation}
  H_I = \sum V_{s_1 s_2 s_3 s_4} \, a^\dag_{s_1} a^\dag_{s_2} a_{s_3} a_{s_4} \, ,
\end{equation}
where the coefficients $V_{s_1 s_2 s_3 s_4}$ are  random  Gaussian variables
with  zero mean and  variance $V^2 = <V_{s_1 s_2 s_3 s_4}^2>$. Thus, the parameter $V$ determines the strength of the interaction between particles. This TBRI-model has been invented long ago \cite{TBRI}  as an improvement of  full random matrix models which do not take into account the two-body nature of the interaction. Indeed, differently  from the well developed random matrix theory (RMT) the TBRI-model is much more physical since it relates the properties of observables in terms of single-particle states, with the properties of many-body spectra and eigenstates. 

\begin{figure}[t]
  \vspace{0.cm}
  \includegraphics[scale=0.6]{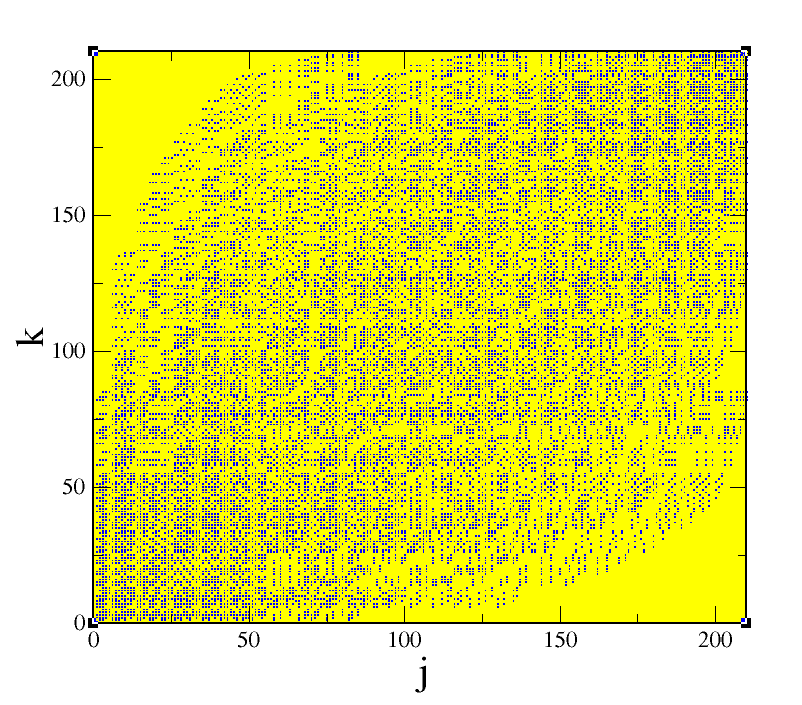}
  \caption{(Color online) Sparsity of off diagonal matrix elements. Each point represents a non-zero matrix element.
Here we consider  $N=4$ particles in $M=7$ levels.
}
  \label{ff1-sparse}
\end{figure}

One of the distinctive properties of the TBRI matrices, in connection with full random matrices, is the presence of many zeros in the structure of the total Hamiltonian $H$, see Fig.\ref{ff1-sparse} for an example. This happens since only a part of the unperturbed many-body states of $H_0$ are directly connected  by the interaction $H_I$. The number of non-zero off-diagonal matrix elements turns out to be small in comparison with the total number $N_H \times \mathcal{N_H}$ of matrix elements of $H$. This fact is important in view of the analysis of the statistical properties of the  eigenstates. Another feature of the TBRI matrices is their band-like structure due to which the many-body states that are close to the ground state are very different from the chaotic eigenstates described by the RMT, as will be discussed in the following.

The size of energy spectrum of $N_H$ can be easily estimated in terms of the model parameters (number of particles, number of single-particle states and  mean level spacing between them). Indeed, the ground state can be obtained by placing all bosons in the lowest single-particle level with the energy $\epsilon_1 \simeq 0$, so that its many-body energy is $E_{low} \simeq 0$. On the other hand, the most excited state is obtained by placing all bosons in the highest $M$-th level $\epsilon_M$ so that its energy is $E_{up} \simeq N M d$.  Since  the number of states is given by $N_H$, one has an average (many-body) level spacing $ D = NMd/N_H$. A naive perturbation theory argument  will give that the typical interaction strength, $V$, should be larger than $D$ in order to  mix many unperturbed energy levels. However, such an approach is not correct due to the high sparsity of the Hamiltonian matrix. Therefore, one should relate the interaction $V$ not to $D$ but to the effective level spacing $d_f$ between directly coupled many-body states.

This quantity $d_f$ depends locally on the unperturbed energy level $E_0^k$ and it can be estimated~\cite{flam} from the relation,
\begin{equation}\label{dfeq}
  d_f (E_0^k) = \frac{(\Delta E_0)_{eff}^k}{M_{eff}^k} \, ,
\end{equation}
where $(\Delta E_0)_{eff}^k$ is the maximal energy range of the unperturbed states effectively coupled to the state with energy $E_0^k$, and $M_{eff}^k$ is the number of those states. Note that in Eq.~(\ref{dfeq}), we put explicitly the dependence of $d_f$ on $E_0^k$.

\begin{figure}[t]
\vspace{0.cm}
\includegraphics[scale=0.7]{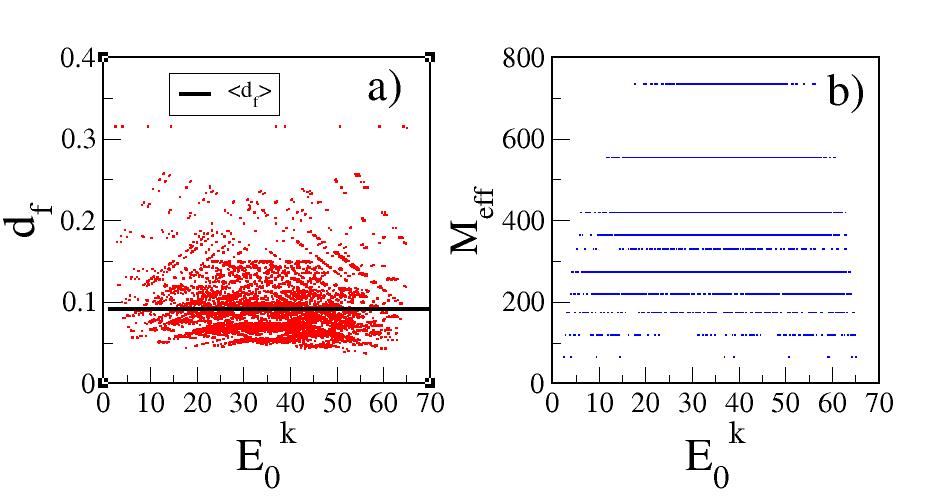}
\caption{(Color online) a) Effective average spacing between many-body states with unperturbed energy $E_0^k$.
b) Number of effectively coupled states for each many-body state with unperturbed energy $E_0^k$.
Here is $N=6, M=11$.
}
\label{ff1a-df}
\end{figure}

Since the Hamiltonian matrix is very sparse, we have $M_{eff}^k \ll N_H$ and $d_f \gg D$. The values of $d_f$ for  $N=6$ and $M=11$ are shown in Fig.\ref{ff1a-df}a). As one can see, the value of $d_f$ is an irregular function of $E_0^k$ and the average cannot be considered as a good approximation for each single many-body state $|k\rangle$ (note that the values of $d_f$ are different for different states $|k\rangle$ and for the parameters given in Fig.\ref{ff1a-df}a) they span roughly one order of magnitude). The number of non-zero off-diagonal elements along any line of the Hamiltonian matrix $H$ is much smaller than $N_H$. For example, in Fig.\ref{ff1a-df}b) the maximal number is $735$, which has to be compared with $N_H = 8008$. Thus, at maximum only $\sim 10 \%$ of the many-body states are coupled, and there are some states which are coupled to only $65$ states (less than $1 \%$ of the available many-body states). This fact suggests that great care should be taken  to get a  correct estimate for $d_f$.

Although there is no analytical estimate of how the number of non-zero off-diagonal elements depends on a specific line in the structure of the Hamiltonian $H$, we can estimate the lower and upper bounds for the number of non-zero matrix elements in the Hamiltonian matrix $H$. Let us first start with the lower bound, ${\cal M}_{min}$. It can be obtained by taking into account that the many-body states having the smallest number of links to other states, can be presented as $| 0,...,0,N,0,...,0\rangle $, which means that all Bose-particles occupy only one single-particle level $s$. Let us consider, for simplicity, only the many-body states directly coupled by the interaction $H_I$. One set of the off-diagonal elements corresponds to the states obtained by  moving only one particle from the $s$-th level occupied by $N$ bosons, to any ($j$)  of the other $M-1$ levels. This corresponds to the action of the operator $a^\dagger_s a^\dagger_j a_s a_s$ and it  gives rise to $M-1$ different coupled states. Another set of the off-diagonal elements corresponds  (for $N>2$) to those states obtained by  moving $2$ particles from the $s$-th level to any other $j$ and $k$ levels (among the $M-1$ available). The corresponding operator is $a^\dagger_j a^\dagger_k a_s a_s$. The number of such coupled states is the same as the number of possible states obtained placing $2$ Bose particles on $ M-1$ levels, i.e $\frac{[2+(M-2)]!}{2!(M-2)!}=M(M-1)/2 $. Summing these two numbers, one gets ${\cal M}_{min} = (M-1)(M+2)/2$.

As for the upper bound ${\cal M}_{max}$, it can be obtained by taking into account that the many-body states having the largest number of connected states can be presented as $| 0,1..,0,1,0,1..,0\rangle$ 
that corresponds to all $N$  particles distributed among the $M >N$ orbitals in order to have a single (or null) occupation in each orbital. These many-body states are coupled to two different sets of many-body states:

A) states created by  one-particle hopping, described by the action of $a^\dagger_s a^\dagger_k a_s a_l$ with $k\ne l$. For each of the $N$ occupied states there are $M-1$ possible ways to move, resulting, in total, in $N(M-1)$ coupled states.

B) states created by  moving  two particles into two levels different from the initial ones
(the corresponding operator is $a^\dagger_s a^\dagger_j a_k a_l$ with $k\ne l$). This action involves $N(N-1)/2$ different pairs of states and for each of them there are many different available levels. The number of such states  is the same as the number of ways to put  2 bosons in $M-2$ different levels, i.e
$
\frac{[2+(M-3)]!}{2!(M-3)!}= (M-1)(M-2)/4
$.
Thus, the number of coupled states is $N(N-1)(M-1)(M-2)/4$.

Summing up A) and B) one gets the upper bound for $M_{eff}$,
\begin{equation}
\label{meff-max}
{\cal M}_{max} = N(M-1) \left[ 1 + \frac{(N-1)((M-2)}{4}  \right]
\end{equation} 
In the particular case of  half-filling for which $M=2N+1$, the estimate for $N_H$ reads,  
$
N_H \sim (\frac{27}{4})^N
$
for large $N$. Therefore, in this limit we have ${\cal M}_{max} \sim N^4$, so that 
${\cal M}_{max}/N_H \to 0$, for $N \to \infty$. This estimate demonstrates that, in general, with an increase of number of bosons, the sparsity of the Hamiltonian matrix strongly increases and one should expect strong deviations from the properties of energy spectra and eigenstates   predicted by the RMT \cite{TBRI}. 

\section{STRENGTH FUNCTION : CROSSOVER FROM BREIT-WIGNER TO GAUSSIAN }

In our setup the eigenstates of the total Hamiltonian $H$ are presented in the basis of $H_0$. The eigenstates of $H_0$ can be obtained creating $N$ particles in $M$ single particle levels, i.e.    $|k\rangle = a^\dagger_{k_1}...a^\dagger_{k_N} |0\rangle$. In what follows for the ({\it unperturbed}) eigenstates of $H_0$ we use the Latin letters, in contrast with the Greek letters which stand for ({\it exact}) eigenstates of $H$. Thus, it is natural to express the exact eigenstates in terms of the unperturbed ones,
\begin{equation}
\ket{\alpha} = \sum_k C_k^{\alpha} \ket{k} ,
\label{ldos}
\end{equation}
where 
\begin{equation}
 H|\alpha \rangle = E^\alpha |\alpha \rangle  ;  \,\,\,\,\,\,\,\,\,\,\, H_0 |k\rangle = E^0_{k} |k\rangle .  
\label{ldos1}
\end{equation}
As one can see, the coefficients $C_k^{\alpha}$ form the structure of exact eigenstates in the unperturbed basis. In order to characterize the eigenstates in dependence on their energy $E$, it is useful to introduce the {\it $F-$function} \cite{flam},
\begin{equation}
F^\alpha (E) = \sum_k |C_k^\alpha |^2 \delta(E-E^0_k),
\end{equation}
which is the energy representation of an eigenstate. From the components $C_k^{\alpha}$ one can construct the {\it strength function} (SF) of a basis state $\ket{k}$,
\begin{equation}
F_k(E) = \sum_{\alpha} |C^{\alpha}_k|^2 \delta (E-E^{\alpha}) \, ,
\label{ld1}
\end{equation}
 also known as  {\it local density of states}. The SF shows how the unperturbed basis state $\ket{k}$ decomposes into the exact eigenstates $\ket{\alpha}$ due to the interaction $H_I$. It can be measured experimentally and it is of great importance since its Fourier transform gives the time evolution of an excitation initially concentrated in the basis state $\ket{k}$ \cite{our2016}. Specifically, it defines the survival probability to find the system at time $t$ in the initial state $\ket{k}$.

\begin{figure}[t]
  \vspace{0.cm}
  \includegraphics[scale=0.7]{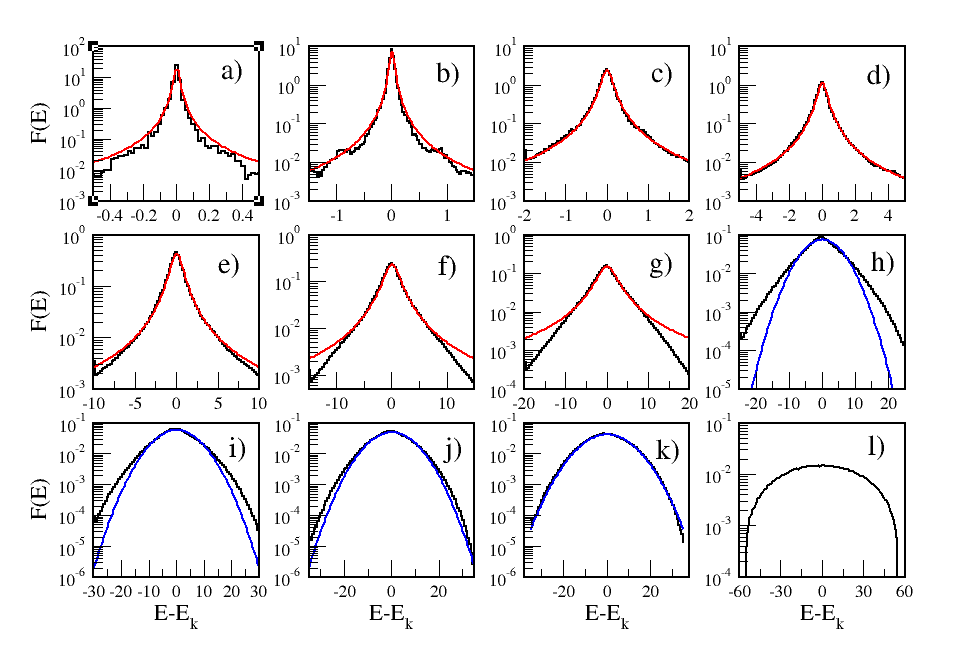}
  \caption{(Color online) The BW-Gaussian transition for the SF as a function of the interaction strength $V$. Red curves are the Breit-Wigner fits while blue curves lines represent the Gaussian fits. The SF have been computed taking $100$ eigenstates in the middle of the energy spectrum for $100$ different realizations of the disorder. From a)...to l), one has $V=0.01, 0.02, 0.04, 0.06, 0.1, 0.14, 0.18, 0.26, 0.31, 0.35, 0.4, \infty$.}
  \label{ff3-ldos}
\end{figure}

On increasing the interaction strength $V$, the SF undergoes three different transitions: i) from a delta-like function (perturbative regime) to the Breit-Wigner (BW) with the width well described by the Fermi golden rule. With a further increase of $V$, the form of the SF tends to a Gaussian. This scenario has been thoroughly studied numerically and observed experimentally \cite{horacio}.
A further increase of the perturbation (when $H_0$ can be neglected) formally corresponds to the situation when the form of the SF is defined by the density of states (DOS) of the perturbation $H_I$ only (see, also, \cite{kota}). Therefore, the maximal size of the SF is defined by the size of the DOS of $H_I$. It should be stressed that in our model  the DOS of both $H_0$ and $H_I$ have a Gaussian shape, a property which strongly simplifies the analytical estimates. Examples of these transitions are shown in Fig.\ref{ff3-ldos}. As one can see, for moderate  values of $V$ the first two transitions are clearly visible. As for the third transition, it has a restricted  physical sense since in this case $H_I$ is much larger than $H_0$ and it cannot be treated as a perturbation. 

To analyze the properties of the BW-Gaussian transition, it is useful to estimate the half-width $\Gamma_E$ due to the Fermi golden rule, 
\begin{equation}
\Gamma_E=2\pi V^2 /d_f \, ,
\label{Gamma}
\end{equation}
together with the square root of the second moment ($\Delta E_k)^2$ of the SF,
\begin{equation}
(\Delta E_k)^2=\sum_{j\ne k} H_{kj}^2
\label{sec_moment}
\end{equation}
Note that this expression for $\Delta E_k$ is exact and  is valid for any value of $V$, in contrast with the half-width $\Gamma_E$ which is derived using perturbation theory. 

\begin{figure}[t]
  \vspace{0.cm}
  \includegraphics[scale=0.6]{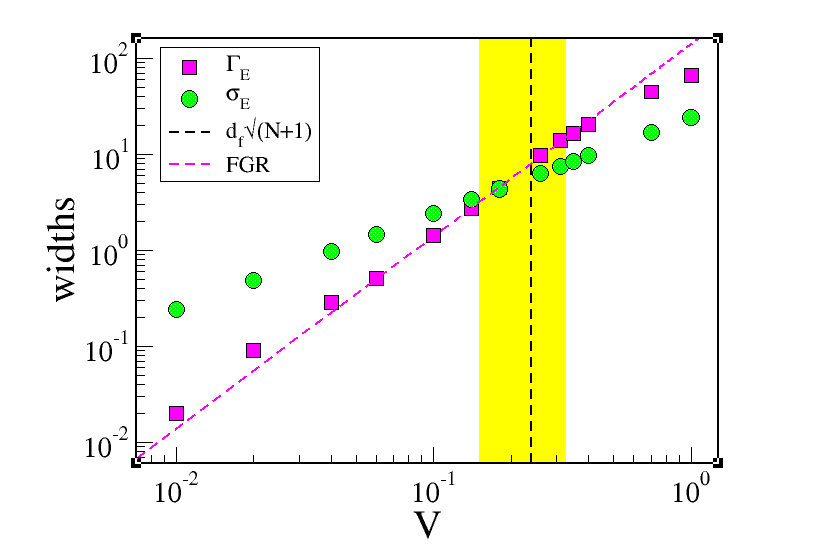}
  \caption{(Color online) The width of SF defined by $\Gamma_E$ and $\Delta E_k$ for the energy $E_k$ taken in the middle of the energy spectrum (see Fig.~\ref{ff3-ldos}). Vertical dashed line corresponds to $V=\langle d_f \rangle \sqrt{N+1}$ and the yellow region is obtained considering one standard deviation of $d_f$. Dashed magenta line is the Fermi golden rule $2\pi V^2/\langle d_f \rangle $ drawn for the reference.}
  \label{ff4-ldos1}
\end{figure}

From the results shown in Fig.~\ref{ff4-ldos1} one can see that while $\Gamma_E$ is quadratic in the perturbation, $\Gamma_E \propto V^2$, the width defined by Eq.~(\ref{sec_moment}) is linear in $V$, $ \Delta E_k \propto V$. Quite remarkably they intersect at the value of $V$ inside the yellow region obtained from the vertical lines $(\langle d_f \rangle \pm\delta d_f)\sqrt{N+1}$ where $\delta d_f$ is the standard deviation and we have taken into account the correction $\sqrt{N+1}$ suggested  in Ref.\cite{luis}.  This is very important; despite the high variation of $d_f$ with the unperturbed energy $E_0^k$ the estimates correctly capture the transition from the BW regime to the Gaussian regime of the SF. As is manifested in Refs.\cite{flam,our2012,our2016}, the Gaussian regime corresponds to  strong quantum chaos characterized by the chaotic structure of many-body eigenstates. 
 
\section{EIGENFUNCTIONS: ONSET OF CHAOS}

\begin{figure}
  \vspace{0.cm}
  \includegraphics[scale=0.55]{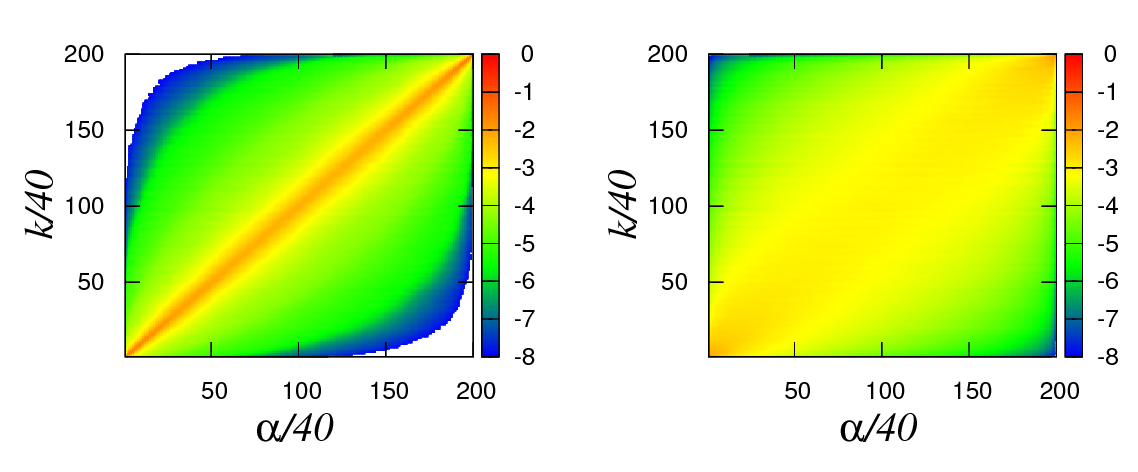} 
  \caption{(Color online) Components of the eigenfunctions, $\log_{10} |C_k^{(\alpha)}|^2$, for two different
perturbation strengths : left $V=0.1$ (the BW regime), right $V=0.4$ (Gaussian regime).}
  \label{ff5-amp}
\end{figure}

In order to demonstrate the onset of quantum chaos in terms of many-body eigenstates, let us concentrate on their global properties. In Fig.\ref{ff5-amp} we show the square values of the amplitudes $|C_k^{(\alpha)}|^2=|\braket{k}{\psi_\alpha}|^2$ of the eigenstates of the total Hamiltonian $H$, as a function of both $\alpha$ and $k$ (suitably renormalized). Two values of the interaction strength have been used, $V=0.1$ (left panel, BW regime) and $V=0.4$ (right panel, Gaussian regime). A clear observation is a pretty nice symmetry ($\alpha \to k$) and the fact that in the BW regime the eigenstates are much more ``localized'' than in the Gaussian regime. Here by the term ``localized'' we mean that the eigenstates occupy only a fraction of the total many-body space. In Fig.\ref{ff5-amp} the regions mostly occupied by eigenstates are presented by the yellow and red colors, see the corresponding scale on the right side  panels in Fig.~\ref{ff5-amp}. These data manifest a quite important symmetry between the exact eigenstates of $H$ in the basis of $H_0$ (vertical lines for fixed $\alpha$) and the unperturbed eigenstates of $H_0$ presented in the basis of $H$. This fact can be effectively used for the analytical estimates of the properties of eigenstates. 

 \begin{figure}[t]
  \vspace{0.cm}
  \includegraphics[scale=0.7]{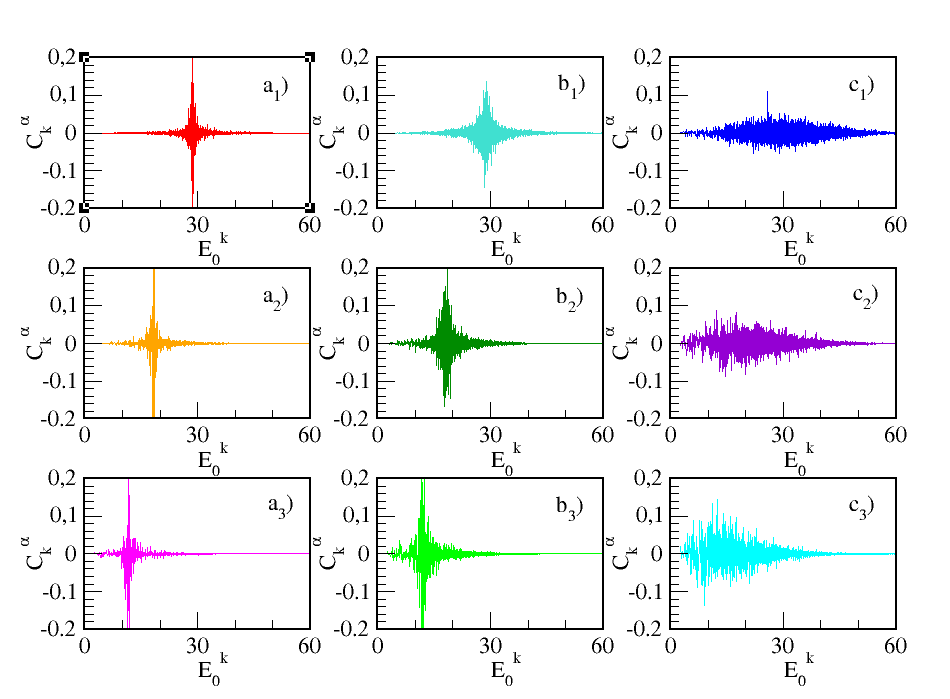}
  \caption{(Color online) Structure of the eigenstates of $H$ in the unperturbed energy representation. 
First column : $V=0.04$, energy $E^\alpha = 28.67 $ (a1), $E^\alpha = 18.17 $ (a2), $E^\alpha = 11.6 $ (a3);
Second column : $V=0.1$, energy $E^\alpha = 28.51 $ (b1), $E^\alpha = 17.88 $ (b2), $E^\alpha = 11.27 $ (b3);
Third column : $V=0.4$, energy $E^\alpha = 25.93 $ (c1), $E^\alpha = 13.24 $ (c2), $E^\alpha = 6.05 $ (c3).
 }
  \label{ff12-eig}
\end{figure}

Another information can be drawn from the structure of the many-body eigenstates in the unperturbed energy representation, as  shown in Fig.~\ref{ff12-eig} for different  energy values $E^\alpha $. One can see that the region occupied by the eigenstates in the unperturbed energy space $E_0^k$ strongly increases with the interaction $V$. However, even when the perturbation is   strong enough to produce a  Gaussian form for the SF (e.g. for $V \simeq 0.4$), the eigenstates do no fill the whole energy space defined by $H_0$. Note that even in the limit case when $H_0$ is neglected (this formally corresponds to the limit $V \rightarrow \infty$), only at the energy band center the eigenstates are almost completely delocalized and can be compared with those predicted by the RMT (however, with a clear deviation from them). When the energy is close to the band edges, (close to the ground state) the eigenstates deviate more and more from the truly Gaussian delocalized states. This fact is quite typical for isolated systems of two-body interacting particles (see also discussion in \cite{our2016}).  

A close inspection of whether the eigenstates can be treated as a superposition of pseudo-random numbers (components $C_k^{(\alpha)}$) can be done when the effective number $N_{pc}$ of principal components is large. The degree of randomness of the components can be numerically checked by proving that the fluctuations of $|C_k^{(\alpha)}|^2$ with respect to their mean values are described by a Gaussian distribution. The corresponding numerical test is very sensitive for determining whether the eigenstates can be indeed considered as chaotic on a finite scale of the unperturbed basis (see also Ref.~\cite{benet}). We did such a test of the randomness for the eigenstates and   only for infinite interaction (i.e. neglecting the mean field part $H_0=0$) the fluctuations of their components with respect to the mean values are, indeed Gaussian. In this test, we have used a large number of the TBRI Hamiltonian matrices with a different choice of random two-body matrix elements. 

\begin{figure}[t]
  \vspace{0.cm}
  \includegraphics[scale=0.55]{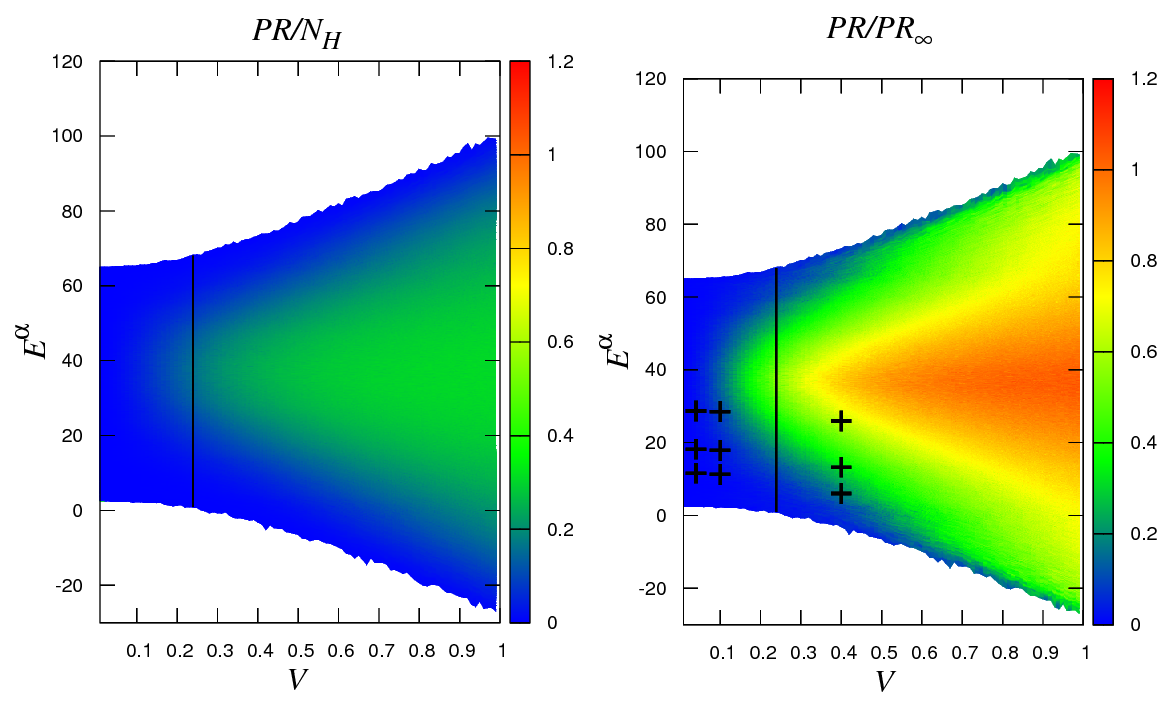}
  \caption{(Color online) Left panel:  Rescaled Participation Ratio as a function of both the energy $E^\alpha$ and the interaction strength $V$. Left panel: the PR rescaled to $N_H$, right panel: the PR rescaled to its maximal value, $PR/PR_\infty$. Crosses represent the rescaled participation ratios   for the eigenstates shown in Fig.~\ref{ff12-eig}. Vertical black lines stand for $V= \langle d_{eff}\rangle \sqrt{N+1}$.
}
  \label{ff7-pr}
\end{figure}

To quantify the degree of localization we use the \emph{Participation Ratio} (PR) of an eigenfunction, defined as 
\begin{equation}\label{PReq}
  PR_\alpha = \frac{1}{\sum_k |C_k^{(\alpha)}|^4} \, ,
\end{equation}
where $\alpha$ labels the many-body eigenstate $\ket{\alpha}$. The degree of localization, compared to the size of the matrix is shown in Fig.\ref{ff7-pr}. The left panel demonstrates that even in the middle of the energy band and for strong interaction $V\simeq 1 $
the eigenstates occupy only a fraction of the total energy region (less than $40\%$). In the right panel we show the participation ratio with respect to its maximal value $PR_\infty$ obtained considering the Hamiltonian with the two-body part $V$ only which
 defines implicitly the available energy shell.
As one can see, for $V > \langle d_f \rangle  \sqrt{N+1}$ (indicated by a black vertical line) a large part of the spectrum (the red one) has a ratio $\sim 1$ which means that eigenstates are mostly delocalized in the energy shell determined by the SF. The concept of  energy shell is very important in  order to  understand that  localization should be considered only with respect to this energy shell and not  to the total energy range (see details in Refs.~\cite{shell,our2012,our2016}. Thus,  thermalization is directly related to the delocalization of many-body eigenstates in the energy shell, although on the scale of the total energy spectrum these eigenstates can be effectively localized. 

\section{SINGLE-PARTICLE OCCUPATION NUMBERS}
In this section we analyze the occupation number distribution  (OND) $n_s^\alpha$  for an energy eigenstate $|\alpha\rangle$, defined as follows:
\begin{equation}
\label{spond}
n_s^\alpha = \langle \alpha | \hat{n}_s | \alpha \rangle = \sum_k |C_k^\alpha|^2
\langle k | \hat{n}_s | k \rangle.
\end{equation}
One can see that the OND (\ref{spond}) depends on two ingredients: the probabilities $|C_k^\alpha|^2$ and the integer numbers $\langle k | \hat{n}_s | k \rangle$ that for the bosons take the values $0,1,2,...N$ depending on how many particles occupy the single-particle level $s$ with respect to the many-body state $| k \rangle $. Above we have seen that under some conditions the eigenstate $| \alpha \rangle$ of $H$ consists of many uncorrelated components. In this case one can substitute $|C_k^\alpha|^2$ in the sum by a smoothed SF. The latter can be obtained either by an average over a number of eigenstates with close energies, or inside an individual eigenstate, for example, with the use of the ``moving window" average \cite{our2016}. 

The key question is: under which conditions the OND is described by the conventional Bose-Einstein distribution (BED) ? This question is far from being trivial since BED has been derived in  statistical physics for non-interacting particles in the thermodynamic limit. Our interest here is in the situation when the number of interacting particles is finite and  relatively small. To address properly this question we start with the basic relations,
\begin{equation}
\sum_s n_s = N
\qquad \sum_s \epsilon_s n_s = E ,
\label{fco}
\end{equation}
where $N$ is the total number of bosons and $E$ is the energy of a system for which  the inter-particle interaction is neglected. As is known, the solution of these equations for $N,M \rightarrow \infty $ leads to the famous BED,
\begin{equation}
n_s^{BE}= \frac{1}{e^{\beta(\epsilon_s-\mu)}-1}
\label{fbe}
\end{equation}
where the constants $\beta$ and $\mu$ are the Lagrange multipliers that can be associated with the inverse temperature and the chemical potential respectively. Below we show that one can speak of  BED even if the system is isolated; moreover, this distribution emerges on the level of a single eigenstate of the total Hamiltonian. Inserting (\ref{fbe}) into (\ref{fco}), one can  obtain both $\beta$ and $z=e^{\beta\mu}$ as a function of  $N$ and $E$. If we further fix the number of particles $N$ we obtain two functions $z(E) $ and $\beta(E)$. Therefore, for any energy $E$ we may have a well defined BED. However, the key question is: what is the influence of the inter-particle interaction if it is relatively strong ? The answer is that one can take into account the inter-particle interaction by properly renormalizing the energy $E$ which stands in the right-hand of Eq.~(\ref{fco}). 

As shown in Refs.~\cite{flam,fausto,our2017}, in order to  describe the actual distributions of $n_s^\alpha $ in terms of a BED one has to substitute the energy $E=E^{\alpha}$ in (\ref{fco}) with the ``dressed'' energy,
\begin{equation}
E^{dres} = \langle \alpha | H_0 | \alpha \rangle \equiv  E^{\alpha} + \Delta_\alpha \, .
\label{she}
\end{equation}
In this way we take into account that the inter-particle interaction changes the effective energy of the system. This energy is higher by some amount $\Delta_\alpha$  than the eigenvalue $E^\alpha$, in the region where the density of states increases, see Ref.~\cite{our2017}. Correspondingly, the effective temperature $T^{dres}$ is higher than that obtained with $E^\alpha$.  

\begin{figure}[t]
  \vspace{0.cm}
  \includegraphics[scale=0.7]{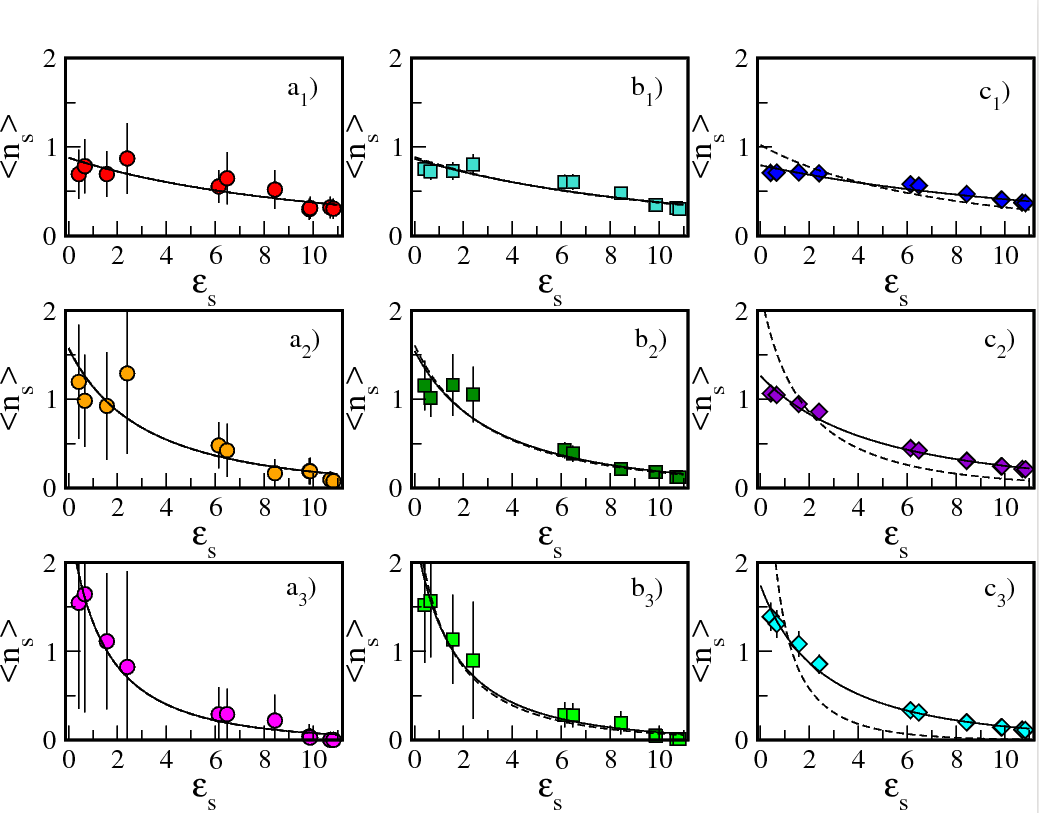}
  \caption{(Color online) Symbols: Average occupation numbers distribution. Full lines : the BE distributions with $E=E^\alpha$. Dashed lines: the BE distribution with $E=E^\alpha + \Delta_{\alpha} $.
First column : $V=0.04$, energy $E^\alpha = 28.67 $ (a1), $E^\alpha = 18.17 $ (a2), $E^\alpha = 11.6 $ (a3);
Second column : $V=0.1$, energy $E^\alpha = 28.51 $ (b1), $E^\alpha = 17.88 $ (b2), $E^\alpha = 11.27 $ (b3);
Third column : $V=0.4$, energy $E^\alpha = 25.93 $ (c1), $E^\alpha = 13.24 $ (c2), $E^\alpha = 6.05 $ (c3).
To get the OND we considered 500 random realizations of the two-body interaction, with an average over $20 $ eigenstates with close energies. In  brackets  $(...)$ we indicated the corresponding eigenstates in Fig.~\ref{ff12-eig}.
 }
  \label{ff9-be}
\end{figure}

With the substitution of $E^{dres}$ in place of $E$ in Eq.~(\ref{fco}) one can get the corrected values of $\mu$ and $\beta$, and from those the corresponding BED which effectively takes into account the inter-particle interaction. Note that the energy shift $\Delta_\alpha$ can be easily calculated numerically for each eigenstate with the use of Eq.~(\ref{fco}). In order to show the effectiveness of this approach we consider the OND in the three regimes, perturbative,  BW and  Gaussian, defined by the form of the SF. The results are summarized in Fig.~\ref{ff9-be} where the nine panels correspond to the crosses in Fig.\ref{ff7-pr}.

According to our results discussed above, for all parameters in Fig.\ref{ff7-pr} the eigenstates are globally localized in the unperturbed basis. However, we already know that only when the inter-particles interaction is strong enough, $V \geq 0.4$ the eigenstates can be treated as chaotic ones, with a large number of principal components. The data demonstrate that in this case the dashed curves which correspond to the OND obtained with no additional shifts $\Delta_\alpha $ are very different with those obtained with the dressed energy (solid curves). And what is very important, these curves perfectly correspond to the actual OND distribution (full symbols in the right panels). In contrast, when the perturbation $V$ is not strong enough  to produce the chaotic states (left and middle column in Fig.~\ref{ff9-be})  dashed and full curves are quite close one to each other. 

Thus, the data clearly demonstrate that for the chaotic eigenstates, the OND obtained with the dressed energy gives a correct description, unlike that one obtained without the shift $\Delta_\alpha $. However, what  can we say for a weaker interaction $V$ ? We can see that both curves, dashed and solid ones, also correspond to the data given by symbols. Can data be taken as the indication that  statistical mechanics works even for non-chaotic eigenstates? The answer is no. The very point is that in non-chaotic case the correspondence between analytical curves and data holds {\it  only in average}. Moreover, the validity of the statistical description requires Gaussian-distributed  fluctuations much smaller than   the mean values. 

 \begin{figure}[t]
  \vspace{0.cm}
  \includegraphics[scale=0.7]{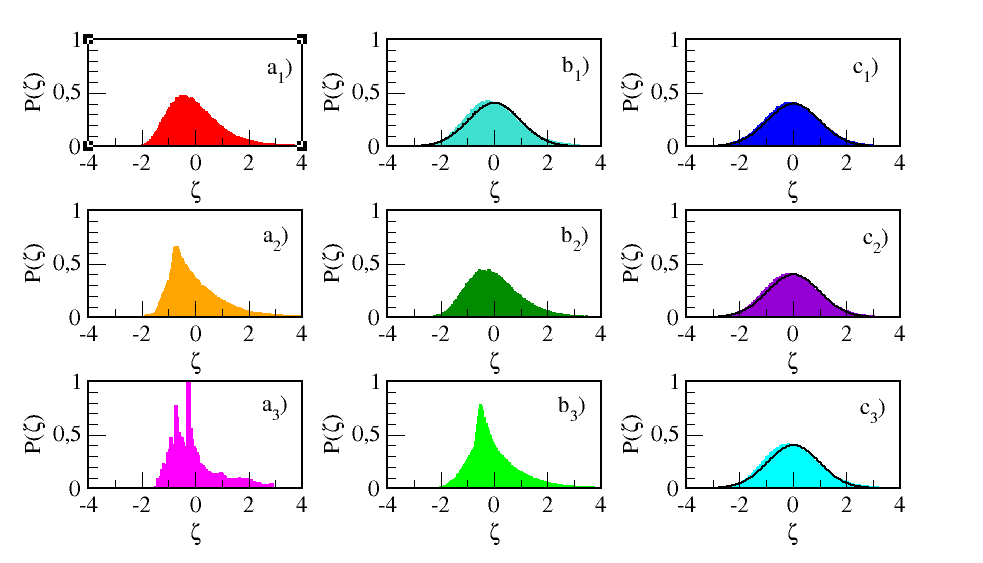}
  \caption{(Color online) Fluctuations of occupation numbers $n_s$ with respect to their mean values for the eigenstates shown in Fig.~\ref{ff12-eig}. Solid curves in $b_1), c_1), c_2), c_3)$  correspond to the Gaussian fits.}
  \label{ff11-be}
\end{figure}  
 \begin{figure}[!h]
  \vspace{0.cm}
  \includegraphics[scale=0.5]{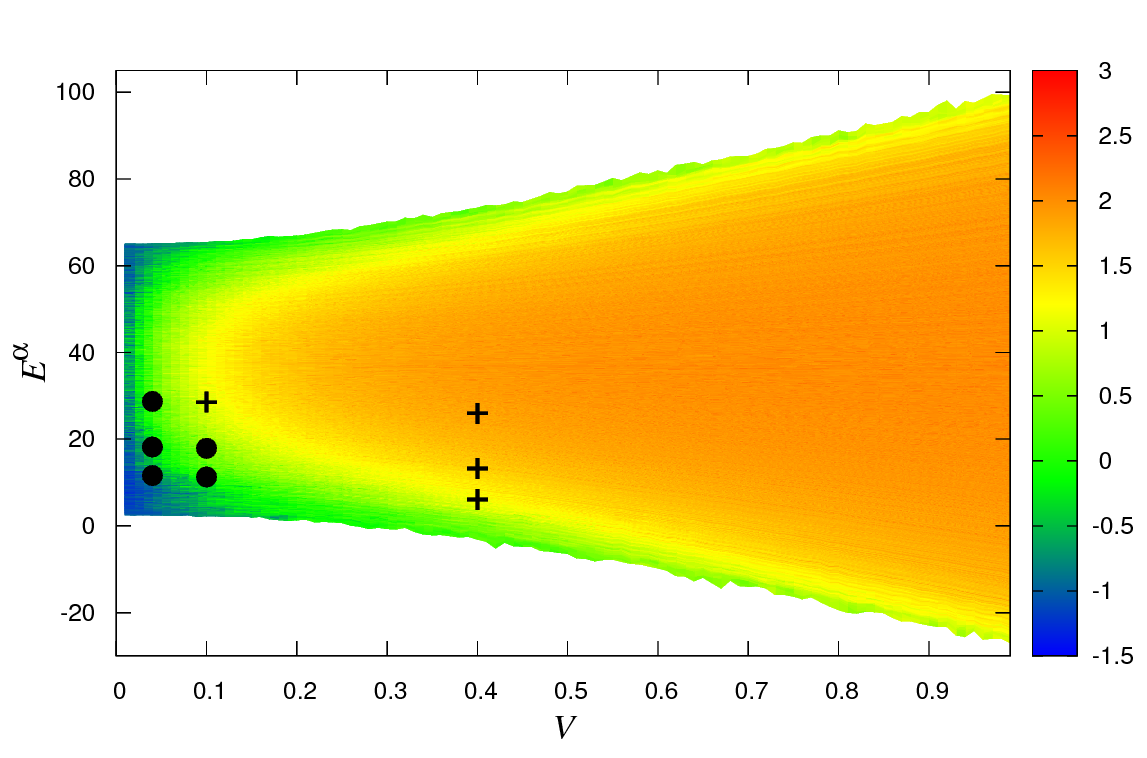}
  \caption{(Color online) $\ln (V/d_{loc})$ as a function of the perturbation strength $V$ and energy $E^\alpha$. Black circles represent non-thermal states, black plusses stand for thermal states.}
  \label{ff10-be}
\end{figure}  

To study the fluctuations of the occupation numbers $n_s$, we introduce the variable
$$
\zeta_s = \frac{n_s - \langle n_s \rangle}{ \delta n_s}
$$
where $\delta n_s^2$ is the variance of $n_s$ corresponding to the single-particle level $\epsilon_s$. 
With this definition the distributions $P(\zeta_s)$ for different $s=1,...,M$ have the same mean ($0$) and variance ($1$) and we can
consider the global distribution $P(\zeta)$ for all orbitals $s=1,...,M$.
 The distributions $P(\zeta)$ for the different eigenstates are reported in Fig.~\ref{ff11-be}. As one can see only 4 cases out of 9 (3 in the right panel and one in the middle one) display   Gaussian  fluctuations (shown by solid curves). We therefore can treat these eigenstates as thermal ones (and they are marked by crosses in Fig.~\ref{ff10-be}).

\section{LOCAL CRITERION FOR THERMALIZATION} 
In the discussion above concerning  the onset of thermalization for chaotic eigenstates, we have introduced  the criterion based on the notion of  mean level spacing $ d_f $ between those many-body states which are directly connected by the two-body interaction. Since in this approach the estimate for $ d_f $ was obtained by an average over all eigenstates, the condition $\langle d_f \rangle \geq V$ should be considered as the global condition for chaos and thermalization. Nevertheless the presence of thermal eigenstates even for $V < \langle d_f \rangle$
(see panel $b_1)$  in Fig~\ref{ff9-be}) suggests
that another condition should be considered as a {\it local criterion} for the onset of thermalization on the level of individual eigenstates. For this reason, let us  introduce the following parameter (dependent on the eigenstate $|\alpha\rangle$) 
\begin{equation}
\label{delta0}
\delta_0^2 = \bra{\alpha} H_0^2 \ket{\alpha} - \bra{\alpha} H_0 \ket{\alpha}^2 \, ,
\end{equation}
which characterizes the energy width occupied by the many-body eigenstate $\ket{\alpha}$ in the unperturbed basis of $H_0$. Together with the number $ N_{pc}$ of principal components occupied by the eigenstate $\ket{\alpha}$ in the unperturbed basis of $H_0$, these two parameters allow to define an effective mean energy spacing, 
$$ d_{loc} = \frac{\delta_0}{N_{pc}}, $$ that the perturbation strength $V$ has to exceed in order to go beyond the perturbative regime. 
The subscript {\it loc} indicates that it depends on the eigenstate energy $E^\alpha$.
Accordingly, in order to have thermal eigenstates we require  $V > d_{loc}$, while for $V < d_{loc}$ we can speak of non-thermal states. Therefore, the region characterized by $V > d_{loc}$ is the ``thermal '' one,  with the temperature  defined via the Bose-Einstein distribution obtained with   the dressed energy.

In Fig.~\ref{ff10-be} we plot $\ln V/d_{loc}$ for all eigenstates with energies $E^\alpha$ in dependence on both the perturbation strength $V$ and $E^\alpha$. The yellow-red region is characterized by a sufficiently strong perturbation; note that this region depends both on $E^\alpha$ and $V$. The symbols in this figure stand for the nine eigenstates considered above. As one can see, only the four black crosses in this region give rise to a  Gaussian distribution of fluctuations.  Therefore, the corresponding eigenstates can be considered as thermal, while the other five (marked by black  circles) are not thermal. It is remarkable that some of these eigenstates have a quite small value of the ratio $PR/N$ (specifically, $PR/N \simeq 0.1$) and also a not a large filling ratio ($\approx 0.3$). These data also manifest that one can treat them as  ``localized thermal''  eigenstates.


\section{ACKNOWLEDGMENTS}
This work was supported by the VIEP-BUAP grant IZF-EXC17-G. The Authors acknowledge fruitful discussions with L.F.~Santos, L.~Celardo, F.~Mattiotti  and R.~Trasarti-Battistoni.


\nocite{*}
\bibliographystyle{aipnum-cp}%

\begin{thebibliography} {99}

\bibitem{reviews} M.~Rigol, V.~Dunjko, and M.~Olshanii, Nature {\bf 452},  854 (2008); A.~Polkovnikov, K.~Sengupta, A.~Silva, and M.~Vengalattore, Rev. Mod. Phys. {\bf 83},  863 (2011); L.~D’Alessio, Y.~Kafri, A.~Polkovnikov, and M.~Rigol, Advances in Physics, {\bf 65},  239-362  (2016); A.~Polkovnikov and D. Sels, Science, {\bf 353}, 752  (2016); D.J.Luitz and Y.B.Lev, Phys. Rev. Lett. {\bf 117}, 170404 (2016); C. Gogolin and J. Eisert, Rep. Prog. Phys. {\bf 79},  056001 (2016). 

\bibitem{our2016} F.~Borgonovi, F.M.~Izrailev, L.F.~Santos, and V.G.~Zelevinsky, Phys. Rep. {\bf 626},  1 (2016).

\bibitem{exp} M.~Greiner, O.~Mandel, T.W.~Hansch, and I.~Bloch, Nature {\bf 419}, (2002) 51; T.~Kinoshita, T.~Wenger, and D.S.~Weiss, Nature {\bf 440},   900 (2006); S.~Trotzky {\it et. al.}, Nature Phys. {\bf 8},  325 (2012); M.~Gring, {\it et al.} Science {\bf 337},    1318 (2012); A.M.~Kaufman et al. Science, {\bf 353}, 794-800 (2016); J.~Smith et al., Nature Phys.  {\bf 12},  907–911 (2016); R.~Nandkishore and D.A.~Huse, Annual Review of Condensed Matter Physics {\bf 6},   15 (2015).

\bibitem{zele} V.~Zelevinsky, M.~Horoi, and B.A.~Brown, Phys. Lett. B {\bf 350},  141 (1995); M.~Horoi, V.~Zelevinsky, and B.A.~Brown, Phys. Rev. Lett. {\bf 74},  5194 (1995); V.~Zelevinsky, B.A.~Brown, M.~Horoi and N.~Frazier,  Phys. Rep. {\bf 276},  85 (1996); V.~Zelevinsky, Ann. Rev. Nucl. Part. Sci. {\bf 46},  237 (1996).

\bibitem{flam} V.~V. Flambaum, A.~A. Gribakina, G.~F. Gribakin, M.~G. Kozlov, Phys. Rev. A {\bf 50},  267 (1994); V.V.~Flambaum, F.M.~Izrailev, and G.~Casati, Phys. Rev. {\bf E 54},  2136 (1996); V.V.~Flambaum, F.M.~Izrailev, Phys. Rev. {\bf E 55},  R13 (1997); V.V.~Flambaum and F.M.~Izrailev, Phys. Rev. E {\bf 56},  5144 (1997).

\bibitem{deutsch} J.M.~Deutsch, Phys. Rev. A {\bf 43},  2046 (1991); J.M. Deutsch, H. Li, and A. Sharma, Phys. Rev. E. {\bf 87},  042135 (2013).


\bibitem{sred} M.~Srednicki, Phys. Rev. E {\bf 50},  888 (1994); J. Phys. A: Math. Gen. {\bf 29},  L75 (1996).


\bibitem{kota} V. K. B.~Kota, A.~Relano,  J.~Retamosa, and M.~Vyas, Journ. of Stat. Mech.: Theory and Experiment  {\bf 2011},  P10028 (2011); N.D.~Chavda, V.K.B.~Kota, V.~Potbhare, Phys. Lett. A, {\bf 376},  2972 (2012).


 
\bibitem{sinai} Ya.G. Sinai, Doklady Akademii Nauk SSSR (in Russian). {\bf 153},  1261  (1963) (in English, Sov. Math Dokl. 4  1818 (1963)); L.A. Bunimovich, Commun Math Phys. {\bf 65},  295 (1979); L.A. Bunimovich and Ya.G. Sinai, Commun Math Phys. {\bf 78},  247 (1980).

\bibitem{fermi} E. Fermi, Phys. Zeit., {\bf 24},  261 (1923); translated into Italian in E. Fermi, Nuovo Cimento, {\bf 26},  105 (1923); {\bf 25},  267 (1923).

\bibitem{FPU} E. Fermi, J. Pasta, and S. Ulam, Los Alamos Sci. Lab. Rep. LA-1940 (1955).

\bibitem{BI05} G.P. Berman and F.M. Izrailev, CHAOS {\bf 15},  015104 (2005).

\bibitem{IC66} F.M. Izrailev and B.V. Chirikov, Dokl. Akad. Nauk SSSR, {\bf 166}, (1966) 57 (in English: Sov. Phys. Dokl. {\bf 11},  30 (1966).

\bibitem{chir1} F.M. Izrailev, A.I. Khisamutdinov, and B.V. Chirikov, Preprint, Inst. of Nucl. Phys., Novosibirsk, USSR Acad. of Sci., (1968) (in English: Los Alamos. Sci. Lab. Rep. LA-4400-TR (1970).

\bibitem{our2012} L.F.~Santos, F.~Borgonovi, and F.M.~Izrailev, Phys. Rev. Lett. {\bf 108},  094102 (2012); Phys. Rev. E. {\bf 85},  036209 (2012); E. J.~Torres-Herrera and L.~F.~Santos, Phys. Rev E, {\bf 88},  042121 (2013); E. J.~Torres-Herrera and L.~F.~Santos, New J. Phys. {\bf 16}, 063010 (2014).

\bibitem{JS85} R.V. Jensen and R. Shankar, Phys. Rev. Lett. {\bf 54}, 1879 (1985).
 
\bibitem{gold} G.~F. Gribakin, A.~A. Gribakina, and V.~V. Flambaum, Aust. J. Phys. {\bf 52},  443 (1999). 


\bibitem{our2017} F.~Borgonovi, F.~Mattiotti, and F.M.~Izrailev, Phys. Rev. E. {\bf 95},  042135 (2017).



\bibitem{TBRI} O.~Bohigas and J.~Flores, Phys. Lett. B {\bf 34},  261 (1971); {\bf 35},  383 (1971); J.B.~French and S.S.M.~Wong, Phys. Lett. B {\bf 33},  449 (1970); {\bf 35},  5 (1971); T.A.~Brody, {\it et al.} Rev. Mod. Phys {\bf 53},  385 (1981); V.K.B.~Kota, Phys. Rep. {\bf 347},  223 (2001).

\bibitem{horacio}  P.R.~Levstein, G.~Usaj, H.M.~Pastawski, The Jour. of Chem. Phys. {\bf 108},  2718 (1998); G.~Usaj, H.M.~Pastawski, P.R.~Levstein, Molecular Physics,  {\bf 95},   1229 (1998).

\bibitem{luis} L.Benet, T.Rupp, and H.A.Weidenmüller, Annals of Physics, {\bf 292},  67 (2001).

\bibitem{benet} L. Benet, F.M. Izrailev, T.H. Seligman, A. Suarez-Moreno, Phys. Lett. A, {\bf 277}, 87 (2000).

\bibitem{shell} G.~Casati, B.V.~Chirikov, I.~Guarneri, F.M.~Izrailev, Phys. Rev. {\bf E 48},  R1613 (1993); Phys. Lett. {\bf A 223}, 430 (1996).

\bibitem{fausto} F.~Borgonovi, I.~Guarneri, and F.M.~Izrailev, Phys. Rev. E. {\bf 57},  5291 (1998); F.~Borgonovi, I.~Guarneri, F.M.~Izrailev, and G.~Casati, Phys. Lett. A {\bf 247},  140 (1998).

\end{thebibliography}

\end{document}